\documentclass[article,nojss]{markjss}

\usepackage{amsmath}
\usepackage{thumbpdf,lmodern}
\usepackage{amssymb}

\allowdisplaybreaks[2]

\author{Mark A. Overton\\Independent Researcher} 
\Plainauthor{Mark A. Overton}

\title{Romu: Fast Nonlinear Pseudo-Random Number Generators Providing High Quality}
\Plaintitle{Romu: Fast Nonlinear Pseudo-Random Number Generators Providing High Quality}
\Shorttitle{Romu: Fast Nonlinear Pseudo-Random Number Generators}

\Abstract{We introduce the Romu family of pseudo-random number generators (PRNGs) which combines
   the nonlinear operation of rotation with the linear operations of multiplication and (optionally) addition.
   Compared to conventional linear-only PRNGs, this mixture of linear and nonlinear operations achieves
   a greater degree of randomness using the same number of arithmetic operations.
   Or equivalently, it achieves the same randomness with fewer operations, resulting in higher speed.
   The statistical properties of these generators are strong, as they pass \pkg{BigCrush} and
   \pkg{PractRand} -- the most stringent test suites available.
   In addition, Romu generators take maximum advantage of instruction-level parallelism in modern superscalar
   processors,
   giving them an output latency of zero clock-cycles when inlined, thus adding \emph{no} delay to an application.
   Scaled-down versions of these generators can be created and tested, enabling one to estimate
   the maximum number of values the full-size generators can supply before their randomness declines,
   ensuring the success of large jobs. Such capacity-estimates are rare for conventional PRNGs.

   A linear PRNG has a single cycle of states of known length comprising almost all possible states.
   However, a Romu generator computes pseudo-random permutations of those states,
   creating multiple cycles with pseudo-random lengths which cannot be determined by theory.
   But the ease of creating state-sizes of 128 or more bits allows
   (1) short cycles to be constrained to vanishingly low probabilities, and
   (2) thousands of parallel streams to be created having infinitesimal probabilities of overlap.
}
\Plainkeywords{PRNG, random number generator, nonlinear}

\Address{
   Mark A. Overton\\
   1026 W. El Norte Pkwy \#56\\
   Escondido, CA 92026\\
   Email: \email{markovertondev@gmail.com}\\
   URL: \url{http://www.romu-random.org}
}

\begin{document}


\section{Introduction} \label{sec:intro}

There are many noncryptological uses of random numbers in software, such as Monte Carlo
simulations, statistical analysis, testing software and hardware, random actions in games, producing
textures and grain in photos, and others. But computers lack an auditable source of true randomness in them,
so random numbers are often obtained from external sources such as the times at which unpredictable
events occur, including packet-arrivals, key-presses, and mouse-movements.
But such sources produce random numbers slowly, so when they are needed quickly, they must be synthesized with
arithmetic. Such an arithmetic algorithm is referred to as a \emph{pseudo-random number generator}
(PRNG). This paper uses the term \emph{generator} when referring to a PRNG.

\citet{MascagniEmpirical} wrote, ``While the quality of a PRNG sequence is extremely important, significant aspects
of a generator's quality are hard to prove mathematically. Therefore, the ultimate test of a
PRNG's quality is empirical''. Accordingly, several test-suites for generators have been developed.
Some well known ones include \pkg{Diehard} \citep{Diehard}, \pkg{Dieharder} \citep{Dieharder}, \pkg{TestU01} \citep{LecuyerTestU01},
and \pkg{PractRand} \citep{PractRand}. \pkg{TestU01} sports three batteries of tests,
among which \pkg{BigCrush} is the most thorough, consuming about $2^{35}$ 32-bit values from the
generator under test and taking four hours. \pkg{PractRand} tests $2^{45}$ bytes by default, making it
even more thorough than \pkg{BigCrush}, taking over three days. \pkg{BigCrush} and \pkg{PractRand} can identify subtle
flaws in a generator, so they were used to test the generators presented in this paper. We
relied most heavily on \pkg{PractRand} (version 0.94) because it provides a measure of goodness of a generator,
and not merely a pass/fail indication. 

This paper presents a new kind of pseudo-random number generator named \emph{Romu}, which stands
for the words \emph{rotate-multiply}, the two primary operations it performs.
The key strengths of generators in the Romu family are:

\begin{itemize}
   \advance\parskip -4pt
   \item Randomness: They have excellent statistical properties, verified by passing the most stringent tests of randomness
         (\pkg{BigCrush} and \pkg{PractRand}).
   \item Speed: They require unusually few instructions and take advantage of instruction-level parallelism (ILP)
         offered by modern superscalar processors, making them fast.
         Their tiny code can be inlined, making them even faster.
   \item Streams: They can generate thousands of independent streams for large parallel jobs.
   \item Assurance: Scaled-down versions can be tested until they fail, providing estimates of the capacities
         of the full-size generators.
\end{itemize}

Our naming convention for these generators appends a suffix to \emph{Romu} indicating how many internal
state-variables a generator has.
\emph{Mono}, \emph{Duo}, \emph{Trio}, and \emph{Quad} signify 1, 2, 3, and~4, respectively.
For example, the state of RomuTrio is held in three 64-bit variables (192 bits).
The name of a Romu generator employing 32-bit arithmetic is also suffixed with \emph{32}.

\begin{figure}
   \centering
   \includegraphics{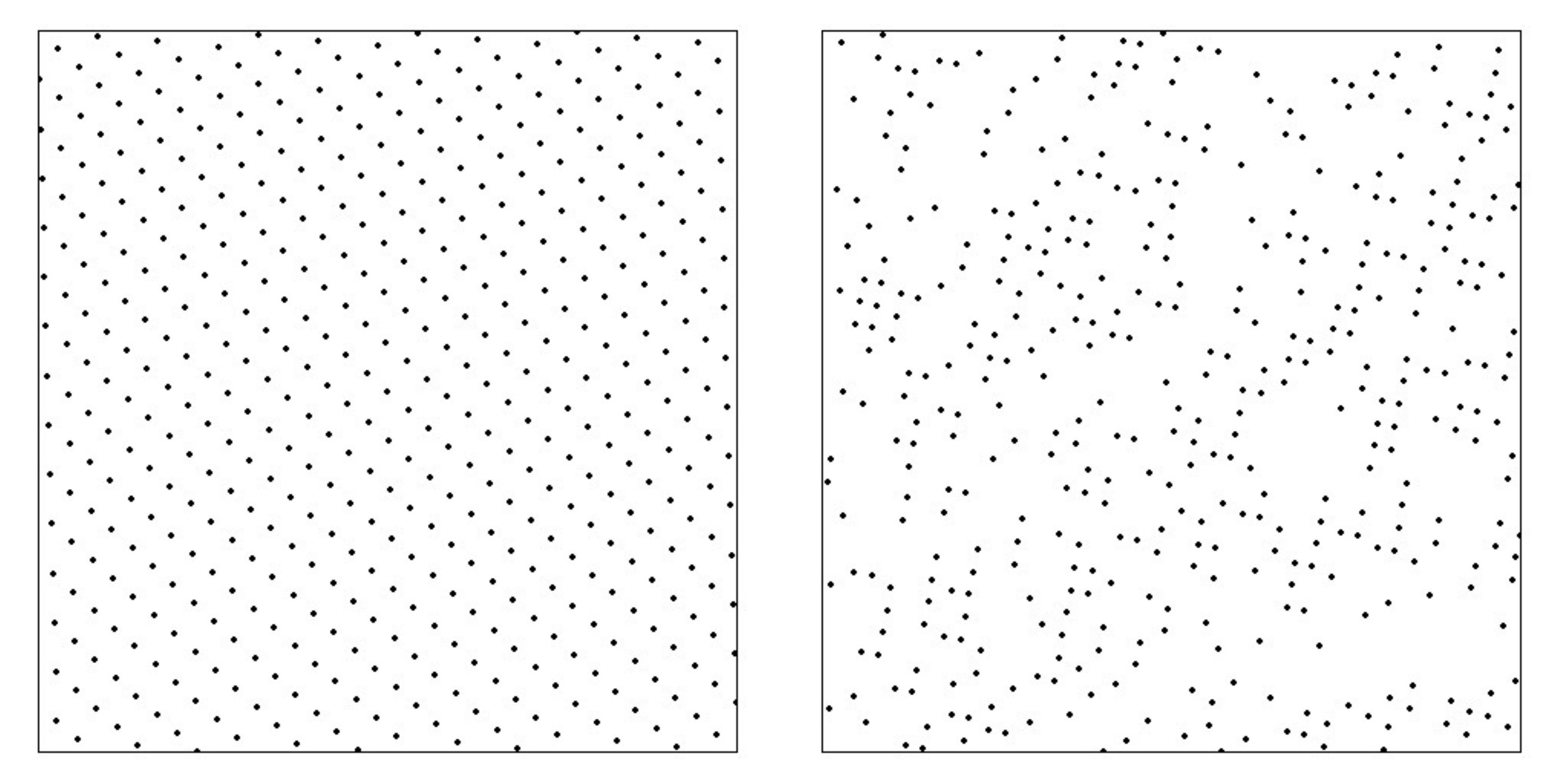}
   \caption{\label{fig:dots} Dot-plots of LCG (left) and Romu (right).}
\end{figure}

To illustrate the high degree of randomness produced by the Romu family, we will compare the
simplest kind of Romu generator with the popular linear congruential generator (LCG) using dot-plots.
The two images in Figure \ref{fig:dots} show all successive pairs of outputs for an entire period of
(1) an LCG with a multiplier of 477, and
(2) a two-operation Romu generator with a rotation of 4 and multiplier of 715.
The arithmetic of both generators was modulo $2^{10}$.
The left plot in Figure \ref{fig:dots} is the usual lattice produced by an LCG.
The right plot produced by the simple Romu generator is clearly more random.

%
%
%

The \proglang{C} source-code of a 64-bit version of the same type of two-operation Romu generator is presented below.
ROTL is a macro or template function that rotates bits leftward.

\begin{Code}
uint32_t result = state;  // state is uint64_t.
state = ROTL(state,32) * 15241094284759029579u;
return result;  // return low 32-bits of 64-bit state.
\end{Code}

This generator, named RomuMono, consists of only \emph{two} assembler-language instructions, making it very fast.
Yet it passes both \pkg{BigCrush} and \pkg{PractRand} (up to $2^{46}$ bytes).
This simple generator reveals the great randomizing power available in a rotation combined with a multiplication.
RomuMono is not discussed further due to its small state-size of 64 bits.

Unlike popular generators such as the LCG, Romu generators do \emph{not}
have a single cycle of states encompassing all possible states.
Instead, they create a pseudo-random permutation of all possible states, and consequently possess multiple cycles
of pseudo-random lengths, including some that are too short. The mathematics of these cycles is intractable, 
so it is not possible to compute the length of the cycle containing the seed.

Fortunately, the probability distribution of cycle lengths is known, and we can rely on such theory
to design generators with infinitesimal probabilities of short cycles.
Furthermore, using a convolution of probability distributions,
we can compute the probabilities of overlap of streams, and thus be
assured that the possibility of either a short cycle or stream-overlap can be safely ignored.


\citet{FogChaotic} proposed the RANROT family of generators which are similar to Romu
in that they are nonlinear with probabilistic cycle lengths.
But his generators employ rotation and \emph{addition},
whereas Romu generators employ rotation and \emph{multiplication}.
Fog rejected multiplication because it was too slow at that time, but we appreciate his insights.

Note that every Romu generator presented in this paper is \emph{not} cryptographically
secure because it is invertible. If an attacker knows its internal state, its prior outputs can
be computed by running the generator backwards, allowing the attacker to decrypt prior messages.


\section{Requirements of Generators} \label{sec:requirements}

In addition to producing good random numbers, we impose several requirements on modern generators.
For large jobs, a generator must have known and sufficient \emph{capacity},
and must provide thousands of \emph{parallel streams} that do not overlap.
To achieve high speed, a generator must make good use of \emph{ILP} and add a minimum of \emph{register pressure}.

\subsection{Parallel streams}

Modern processors contain multiple cores, each of which can run a program or thread independently
of the other cores. High-end computers contain hundreds of processors and several thousand cores.
To take advantage of this available parallelism, large simulations can be divided into sub-jobs to
be run in parallel, one per core. Such parallel computation requires that a generator supply an
independent stream of values for each of the many cores.
There are two popular methods of creating streams: cycle-splitting and parameterizing.

\subsubsection{Cycle-splitting} \label{sec:cyclesplitting}

\citet[p.~2]{MascagniEmpirical} mention and dismiss the
``naive'' method of creating streams in which ``users randomly select seeds on each processor
and hope that the seeds will take them to widely separated portions of the sequence, so that
there will be no overlap between the sub-sequences used by different processors.''
However, for a Romu generator with at least 128 bits of state, we claim that this naive approach is excellent.
Table~\ref{tab:convOverlap} tells us that such a generator can provide tens of thousands (and usually far more)
sub-sequences (streams) with an infinitesimal chance of any overlap occurring among them.
Cycle-splitting is well-suited for Romu generators.

Some generators offer a jump-function allowing one to jump ahead in the cycle by some large number of values.
This method of cycle-splitting results in \emph{no} chance of overlap.

\subsubsection{Parameterizing}

A second method of providing multiple, independent streams is to use a different multiplier
or rotation for each stream produced by a generator.
A severe disadvantage of creating streams by changing multipliers or other constants is that
it is not practical to test every combination of constants. \pkg{PractRand} requires at least
three days to test a generator up to $2^{45}$ bytes, making it impractical to test thousands of
generators. Thus, when employing parameterization, the generator will be \emph{untested},
creating an unknown risk to the job. With at least 128 bits of state, cycle-splitting virtually
guarantees separation of streams in a \emph{tested} generator,
suggesting that cycle-splitting is the safer approach.

\subsection{Instruction-level parallelism}

Good generators rely on instruction-level parallelism (ILP) in the processor to attain high speed.
Modern PC processors can typically execute up to four instructions in one clock cycle, so they are known
as \emph{four-issue superscalar} designs. A clock cycle can be regarded as having four slots in which
an instruction can issued (executed). In each clock cycle, while instructions are being executed,
the processor looks ahead in the instruction stream,
analyzes data-dependencies among upcoming instructions, and fills as many slots as possible
with instructions whose input-values will be known in the next clock cycle.

ILP is illustrated with the following RomuTrio generator which is presented in detail later.

\begin{Code}
uint64_t romuTrio_random () {
   uint64_t xp = xState, yp = yState, zp = zState;
   xState = 15241094284759029579u * zp;
   yState = yp - xp;  yState = ROTL(yState,12);
   zState = zp - yp;  zState = ROTL(zState,44);
   return xp;
}
\end{Code}

This generator consists of five arithmetic operations: a multiplication, two rotations, and two
subtractions. Thanks to ILP, these five instructions consume only three clock cycles as shown
in the following ILP table.

\begin{center}
   \begin{tabular}{cccccc}  \hline
      \rule{0pt}{2.4ex}Cycle &  Slot 1    & Slot 2     & Slot 3               &  Slot 4               &  Multiplier  \\  \hline
      \rule{0pt}{2.5ex}1     & subtract x & subtract y & multiply             &  \phantom{subtract x} &      *       \\
                       2     & rotate y   & rotate z   & \phantom{subtract x} &                       &      *       \\
                       3     &            &            &                      &                       &      *       \\  \hline
   \end{tabular}  
\end{center}

The \emph{Multiplier} column of an ILP table shows when the processor's multiplier is busy. A single
multiplication consumes three cycles in today's Intel processors, which is why most Romu generators presented in this paper
consume three cycles. If multiplication drops to two cycles in future processors, the speed of some Romu generators
will increase proportionately.

A core in an Intel PC CPU has only one multiplier, but it supports overlapping (termed
\emph{pipelining}), so the core can start exactly one new multiplication in each clock cycle. Thus,
up to three multiplications can be active in a given cycle. A constraint in Intel processors
is that at most two shifts and/or rotations can be performed in one clock cycle.

In the RomuTrio generator, there are three sequences of independent computations: a multiplication and
two subtract-rotate pairs. Because these sequences do not depend on each other, each sequence is
executed concurrently as seen in the above ILP table, greatly increasing performance.

Timing measurements confirm this understanding of ILP. Table~\ref{tab:timings} shows the execution times of
combinations of one, two, and three simple RomuMono32 generators (presented later) as measured on an Intel Xeon E5-2643 v3 (3.4-3.7 GHz),
which is the type of processor most commonly seen in workstations used by engineers and scientists. The table
shows the number of milliseconds and clock cycles consumed by $5 \!\times\! 10^9$ invocations of the combination generator.
Each base generator consists of only two instructions (rotate, multiply), and theoretically consumes 4 clock cycles.
(Most generators presented in this paper are structured to consume only 3 clock cycles.)
The source-code of the combination generator in row~2 of the table is shown below.

\begin{Code}
x = ROTL(x,r1) * c1;       // generator 1
y *= c2;  y = ROTL(y,r2);  // generator 2
\end{Code}

Note that the two instructions in the second generator are
reversed (multiply, then rotate) to take advantage of pipelining in the multiplier.

\begin{table}[ht!]
   \centering
   \begin{tabular}{clcc}  \hline
      \rule{0pt}{2.4ex}Row & \phantom{xxx}Generator    &  mSec  & Cycles \\  \hline
      \rule{0pt}{2.4ex}1   & One RomuMono32                &  6238  &  4.6 \\
                       2   & Two RomuMono32 Combo          &  6226  &  4.6 \\
                       3   & Three RomuMono32 Combo        &  6562  &  4.9 \\  \hline
   \end{tabular}
   \caption{\label{tab:timings} Execution times.}
\end{table}

Rows 1 and 2 of Table~\ref{tab:timings} show that two Romu generators in combination consumed
the same time as one (within measurement error), revealing that ILP achieved 100\% parallelism.
Row~3 reveals that even three generators achieved nearly 100\% parallelism.
When used well, ILP boosts performance severalfold.

\subsection{Low register pressure}

The term \emph{register pressure} refers to the number of registers the compiler wants to use at a specific point in
a program. A random number generator is often called from an inner loop, and
to improve performance, the generator can be inlined. Consequently, all registers
devoted to the generator add to the register pressure inside the loop, and if it exceeds the
number of registers in the processor, some variables in the loop will be stored in memory instead,
referred to as \emph{spilling}, reducing speed. To increase ILP, the designer of a generator might arrange
for a calculation to be independent of the others, which will require more registers, causing more
spills and reducing speed, defeating the purpose of increasing ILP. Thus, a compromise must be
made between ILP and register pressure, as both affect the speed of the application.

We regard the usage of up to 5 registers by a generator to be good, 6 as borderline, and 7 or
more as excessive. Most of the Romu generators presented in this paper are acceptable,
as they require 4 to 6 registers when inlined in a loop. The xoshiro256++ generator \citep{xofamily} requires 7-8
registers because it keeps 4 state-variables in registers and its high level of ILP consumes 3
or 4 more. The 128-to-64 bit variant of the pcg-xsh-rs generator \citep{pcg} devours 10 registers
because its 128-bit operations require twice as many registers to hold operands.

The best ways to minimize register pressure are (1) minimize the number of variables in which
the generator's state is stored, and (2) do not allow such a variable to exceed the processor's
word-size (usually 64 bits). For best performance, we suggest avoiding a generator containing
over four state-variables or performing 128-bit or wider arithmetic.

\subsection{Known capacity}  \label{sec:capacity}

The \emph{capacity} of a generator is the quantity of data it can produce before statistical flaws
become detectable. Our testing of multiple types of generators revealed that
a generator's capacity is almost always less than its period -- often far less.
Therefore, 
a user is wise to require that a generator's capacity be measured or estimated.


However, one must keep in mind that a generator's period is usually more important than its capacity
because exceeding its period has worse consequences.
Exceeding its capacity causes subtle correlations to become detectable in the large
quantity pseudo-random data, which will cause little to no harm to many kinds of jobs.
But exceeding its period causes previously-supplied data to be supplied again,
which is a severe correlation that can significantly distort results.


Capacity is not the same as \emph{overtaxing}, a concept introduced by \citet{pcg}.
Overtaxing refers to exceeding a generator's period, and the solution is to increase its period.
But increasing its period will not increase a generator's capacity,
although increasing its state-size can do so.

\begin{figure}[thb]
   \centering
   \includegraphics[width=0.70\linewidth]{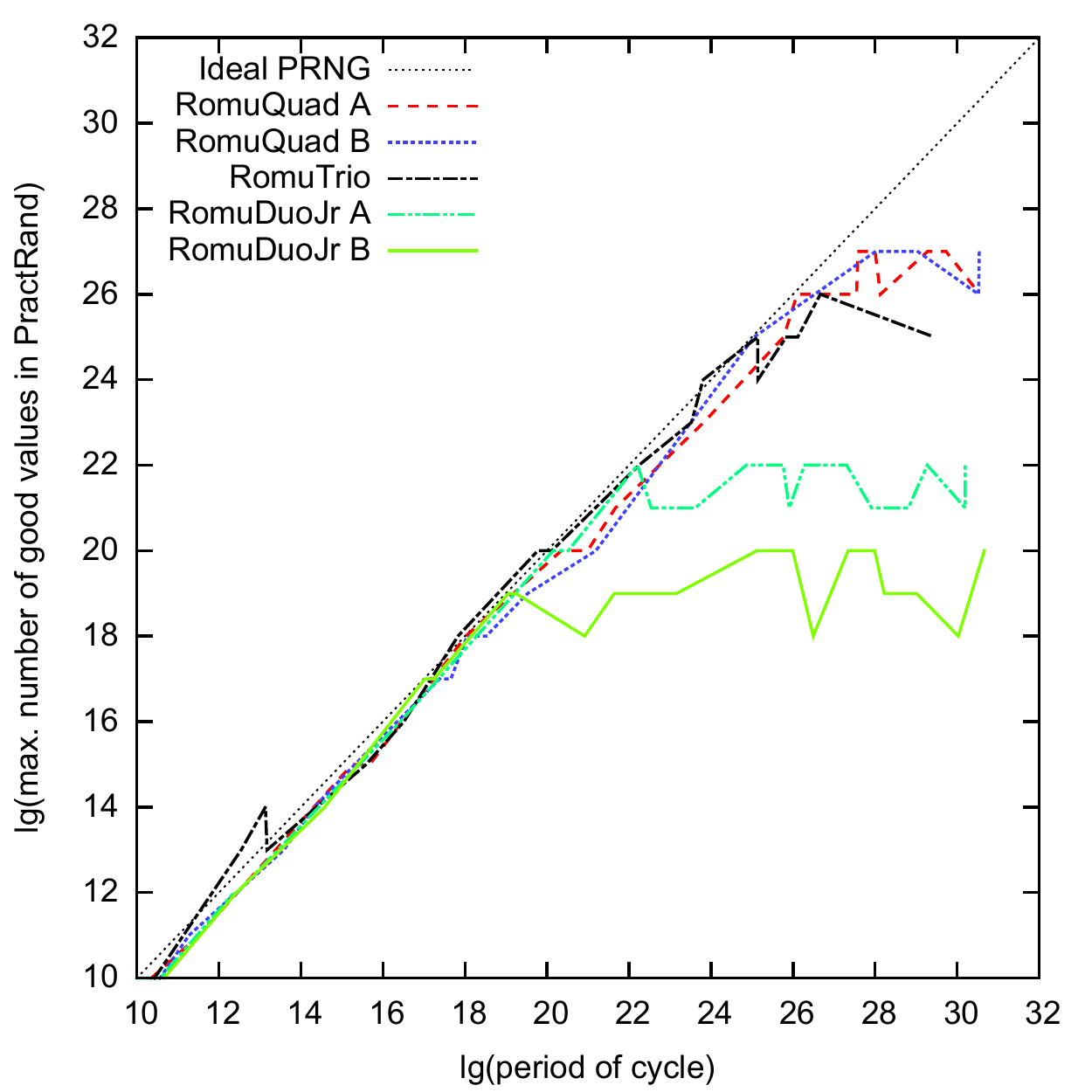}
   \caption{\label{fig:goodValues} Good values versus cycle-period for Romu generators.}
\end{figure}

The existence of multiple cycles (with various periods) in a Romu generator allows us
to graph its capacity as a function of period.
Scaled-down versions of the Romu generators presented later in this paper were created such
that their state size was 32 bits (30 bits for RomuTrio). The generators RomuQuad, RomuTrio,
and RomuDuoJr were selected for this investigation. RomuQuad is the best generator we present and
RomuDuoJr is the worst, consisting of only three arithmetic instructions.
The small 32-bit size made it possible to traverse all of their cycles,
which is not possible with larger state-sizes, and to run \pkg{PractRand} on each of those cycles
and record the number of values output just before the generator began exhibiting anomalous behavior.
The results are plotted in Figure~\ref{fig:goodValues}.

The B-versions in the graph used different multipliers than the A-versions.
The horizontal axis is the base-2 logarithm of period, and the vertical axis is the base-2
logarithm of the maximum number of nonanomalous values (not bytes) supplied by the generators.

The graph reveals that each of these scaled-down generators worked well to a point, and then
leveled onto a bumpy plateau in which continuing to boost period (by selecting longer cycles)
failed to boost the number of good values. The plateau is the generator's capacity.
The arithmetic in RomuQuad is more complex (6~operations) than that in RomuDuoJr
(3~operations). The graph shows that RomuQuad has greater capacity, as would be expected.

An important observation is that before exceeding their capacities, all Romu
generators stayed close to the theoretical ideal generator identified by the dashed 45-degree
line. That is, up to a period of $2^p$ (which must not exceed capacity), each generator could
output close to $2^p$ values before \pkg{PractRand} could detect statistical flaws.
This behavior tells us several things:

\begin{itemize}
   \item Up to its capacity, arithmetic employing a multiplication with rotation(s) produces
         excellent randomness.
   \item With Romu generators, one need not follow the LCG-based rule of thumb that the number of
         values used should not exceed the square-root of the generator's period.
         Figure~\ref{fig:goodValues} tells us that most of the period is usable (up to the generator's capacity).
   \item If a Romu generator happens to be in a small cycle, it will perform nearly ideally.
         Thus, if a job (or stream) requires $2^k$ values, the generator will perform well if the period
         is not under $2^{k+1}$. The 1 was added as a safety margin, and $2^{k+1}$ was assumed not to exceed capacity.

         As an example using numbers for an impossibly huge job, if a generator's state size $s$ is 192 bits,
         and $k\!+\!1$ is 66, then equation \eqref{eq:probCycleLen} tells us that
         the probability that the period will be below $2^{66}$ is $2^{66-192}$ = $2^{-126}$,
         which is the probability of randomly selecting a given grain of sand in all the beaches of the world \emph{twice in succession}.
         This is incomprehensibly tiny, assuring us that such a generator will always perform well.
\end{itemize}

Capacities of Romu generators cannot be determined by testing, as doing so would require centuries.
Therefore, they were estimated based on tests of half-scale, quarter-scale, and eighth-scale generators.
For generators using 64-bit arithmetic, the state-variables in these scaled-down generators were 32, 16, and 8 bits each.
Some of these mini-generators were not the same as those used to produce Figure~\ref{fig:goodValues},
as they were not restricted to 32 bits of state.
Due to their smaller state-sizes, it was possible to run these mini-generators until they failed.
The number of good values a mini-generator produced in a test was its capacity.

Estimated capacities for the corresponding full-size generators were extrapolations
of the mini-generator capacities, requiring one to three doublings of the sizes of the mini-generators.
For each such doubling, the base-2 logarithm of the tested capacity was multiplied by $1.4$.
To be conservative, $1.4$ was used in lieu of $2.0$ because it was observed that
doubling the number of state bits in a generator (by doubling its word-size) multiplies that
logarithm by a factor of $1.4$ to $1.7$, which is less than the $2.0$ suggested by intuition.

\section{Theory of Romu Generators} \label{sec:theory}

The values produced by linear generators consist of one cycle of known length.
These generators are mathematically tractable, and thus much is known about them.
The LCG and MCG are the most widely known linear generators.
Xorshift \citep{MarsagliaXorshift} and the generators from \citet{xofamily} are more recent examples.
But such generators create linear artifacts (such as a lattice structure) in their output values, a
serious problem that greatly reduces the usable portion of their periods \citep{LecuyerReduction}.
\citet[p.~21]{Gentle} recommends not exceeding the square-root of a generator's period, representing a
severe reduction. However, it is important to note that the Blackman-Vigna generators, the LCG-based PCG \citep{pcg},
and enhancements to xorshift perform a post facto scrambling step on the output value (not affecting state),
eliminating their linear artifacts and greatly increasing their capacities.

Romu is nonlinear in the field of integers modulo $2^w$ ($\mathbf{Z}/2^w\mathbf{Z}$), where $w$
is the processor's word-size. Romu's multiplication (and any additions) is linear in this field,
but its rotation is not, resulting in a nonlinear function which is mathematically intractable.
Mathematical analysis of a \emph{linear} generator provides several benefits:
\pagebreak  
\begin{enumerate}
   \advance\parskip -4pt
   \item The generator's period can be computed.
   \item A jump formula can be derived, from which a jump-ahead function can be coded.
   \item In some cases, lattice spacings can be computed, providing measures of randomness.
\end{enumerate}

On the other hand, the benefit of \emph{nonlinear} generators is their freedom from linear artifacts, thus improving
randomness and eliminating the need for a scrambling step, increasing speed. But without the above benefits of analysis,
their periods are unknown, jump-ahead is impossible, and their randomness must be gauged solely by testing.
Also, as mentioned above, Romu generators have multiple cycles of differing lengths.
For this reason, the period obtained by an arbitrary seed is unpredictable, but is probabilistic
as will be seen next.


\subsection{Multiple cycles with probabilistic periods}

The integers in set $S = \{1, \ldots, 2^{s}\!-\!1\}$ are all of the allowable states of a Romu generator
having $s$ bits of state. Note that the all-zeros state is excluded.
Every Romu generator is a bijection because it maps every member of $S$ to every member of $S$.
Consequently, a Romu generator is invertible: Given its state, its prior state can be computed.
Also, the mapping is a permutation of the members of $S$,
meaning the generator rearranges all integers in $\{1, \ldots, 2^{s}\!-\!1\}$.
Crucially, this permutation by a Romu generator is pseudo-random by virtue of its nonlinear arithmetic.
Random permutations consist of multiple cycles of random lengths \citep[pp~160-161]{Knuthvol1},
and every state is in some cycle.
\citet[eq.~8]{FogChaotic} and \citet[p.~518 ex.~17]{Knuthvol1} have disclosed that,
given a random permutation of the integers \{1, \ldots , $n$\},
the probability that a value $x$ is in a cycle of length $m$ is $1/n$.
Curiously, this probability is independent of $m$.
We can verify this counterintuitive statement with the following derivation:
\begin{align*}
   \mathsf{P}(|\text{cycle containing } x| = m) & = \frac{n\!-\!1}{n} \cdot \frac{n\!-\!2}{n\!-\!1} \cdot \frac{n\!-\!3}{n\!-\!2} \cdots \frac{n\!-\!m\!+\!2}{n\!-\!m\!+\!3} \cdot \frac{n\!-\!m\!+\!1}{n\!-\!m\!+\!2} \cdot \frac{1}{n\!-\!m\!+\!1} \\
   & = \frac{1}{n} .
\end{align*}

Thus, the probability that the length of the cycle is less than or equal to $m$ is simply
\begin{align}
   \mathsf{P}(|\text{cycle containing } x| \leq m) & = \sum_{i=1}^{m} \frac{1}{n} \notag \\
   \label{eq:movern}                         & = \frac{m}{n} .
\end{align}

This equation is also presented in \citet[eq.~15]{FogChaotic}.

For any state $x$ (i.e., the generator's seed) that was randomly selected from the $2^s-1$ possible states in $S$,
we want to know the probability that the length of the cycle containing $x$ is less than or equal to $2^k$.
We substitute $m=2^k$ and $n=2^s\!-\!1$ into \eqref{eq:movern}, yielding
\begin{align}
   \mathsf{P}(|\text{cycle containing } x| \leq 2^k) & = {2^k} / {(2^s\!-\!1)}  \notag  \\
   \label{eq:probCycleLen} & \approx 2^{k-s} .
\end{align}

\begin{table}[ht!]
   \centering
   \begin{tabular}{cccc}  \hline
      \rule{0pt}{2.4ex}Row & State size (bits) & Min. cycle len  & Prob. of shorter cycle \\  \hline
      \rule{0pt}{2.6ex}1   &  256              &  $2^{56}$       &  $2^{-200}$ \\
                       2   &  192              &  $2^{56}$       &  $2^{-136}$ \\
                       3   &  128              &  $2^{56}$       &  $2^{-72\phantom{0}}$ \\
                       4   &  \phantom{0}96    &  $2^{56}$       &  $2^{-40\phantom{0}}$ \\
                       5   &  \phantom{0}64    &  $2^{53}$       &  $2^{-11\phantom{0}}$ \\  \hline
   \end{tabular}
   \caption{\label{tab:cycleProbs} Probabilities of shorter cycles.}
\end{table}

Table~\ref{tab:cycleProbs} provides shorter-cycle probabilities for several state sizes based on \eqref{eq:probCycleLen}.
Except in its last row, this table assumes a minimum cycle length of $2^{56}$, which exceeds the longest feasible stream
as it would take over two years to consume $2^{56}$ values at the optimistic rate of one per nanosecond.
The probabilities of encountering cycles shorter than $2^{56}$ are inconceivably low
for state sizes of 256, 192, and 128 bits (rows~1-3). With 96 bits of state (row~4),
the probability for that huge job is one in a trillion. But with only 64 bits (row~5),
a large job of $2^{53}$ values has a shorter-cycle probability of about one in 2000, showing us that 64 bits
of state is not quite adequate for general purpose work when using an invertible nonlinear generator such as Romu.

Based on \eqref{eq:probCycleLen}, we conclude that when the number of bits $s$ in a generator's state
is at least 128, the probability of encountering a cycle shorter than the job or stream is infinitesimal.

\citet{FogChaotic} suggested adding a \emph{self-test} to his RANROT generators
which would detect the completion of a cycle by comparing the state with the seed,
and take an unspecified remedial action. But with a state size of at least 128 bits,
the probability of seeing a too-short cycle is so low that such a test would not be beneficial.

\subsection{Overlaps among sequences}

We must ask ourselves, ``What is the purpose of a seed from the user's perspective?'' The user
selects a different seed in order to obtain a different pseudo-random sequence. 
Also, in Section~\ref{sec:cyclesplitting}, it was recommended that multiple streams be implemented
with cycle-splitting, by selecting streams with seeds.
Thus, the user regards a seed as a sequence-selector, and will be angered if the sequences or streams
resulting from two seeds overlap. Hence, we need to know the probability of overlap.


\subsubsection{Generators with known periods}


Let us define
$l$ as the length of each sequence (or stream) used by an application,
$n$ as the number of such sequences needed, and
$p$ as the (known) period of a generator or that of a specific cycle in a Romu generator.

Let us regard the $n$ sequences as having been generated sequentially (i.e., not in parallel).
The probability that sequence $i$ overlaps a prior sequence is $2il/p$, $i \in [0,n-1]$.
Why the factor of 2? If we think of starting a new sequence $B$ at a random place in the period,
then if its random starting position is within $l$ of the beginning of some prior sequence $A$,
then $B$ will overlap $A$. Hence, each sequence is preceded by a vulnerable zone of length
$l$. A sequence itself is also vulnerable to overlap, so the combined vulnerable block has length $2l$.

To compute the probability that any sequence overlaps another,
we must analyze the two cases of $n$ being even and $n$ being odd.

\begin{align}
   \label{eq:evenover1} \mathsf{P}(\text{overlap | }n\text{  is even}) & = 1 - \prod_{i=0}^{n-1}\left (1 - \frac{2il}{p} \right ) \\
   \label{eq:evenover2}    & = 1 - \prod_{i=0}^{(n/2)-1} \left (1 - \frac{2il}{p} \right ) \left (1 - \frac{2(n-1-i)l}{p} \right ) \\
   \label{eq:evenover3}    & = 1 - \prod_{i=0}^{(n/2)-1} \left (1-\frac{2l(n-1)}{p} + \underbrace{ 4(n-1-i)i \left (\frac{l}{p} \right )^{2} }_\text{insignificant} \right ) \\
   \label{eq:evenover4}    & \approx 1 - \left (1 - \frac{2l(n-1)}{p} \right )^{n/2} \\
   \label{eq:evenprob}     & \approx \frac{(n-1)nl}{p} \\[3ex]
   \label{eq:oddover1} \mathsf{P}(\text{overlap | }n\text{  is odd}) & = 1 - \prod_{i=1}^{n-1}\left (1 - \frac{2il}{p} \right ) \\
   \label{eq:oddover2}     & = 1 - \prod_{i=1}^{(n-1)/2} \left (1 - \frac{2il}{p} \right ) \left (1 - \frac{2(n-i)l}{p} \right ) \\
   \label{eq:oddover3}     & = 1 - \prod_{i=1}^{(n-1)/2} \left (1-\frac{2nl}{p} + \underbrace{ 4(n-i)i \left (\frac{l}{p} \right )^{2} }_\text{insignificant} \right ) \\
   \label{eq:oddover4}     & \approx 1 - \left (1 - \frac{2nl}{p} \right )^{(n-1)/2} \\
   \label{eq:probOverlap}  & \approx \frac{(n-1)nl}{p}
\end{align}

Equations~\eqref{eq:evenover2} and \eqref{eq:oddover2} group the terms in the products into pairs.
Equations~\eqref{eq:evenover3} and \eqref{eq:oddover3} reveal that these pairs
are approximately equal because the expression marked ``insignificant'' is
O$((nl/p)^2)$, making it much smaller than the other terms in the sum when $nl/p \ll 1$.
Equations \eqref{eq:evenover4} and \eqref{eq:oddover4} were simplified using the binomial approximation:
\begin{equation*}
   (1 + x)^n \approx 1 + nx \text{, \quad $|x| \ll 1$} .
\end{equation*}

Like the first approximation above, the binomial approximation adds an insignificant error of O$((nl/p)^2)$.
Both of these approximations increase the final result, making it an upper bound.
And both make small changes in the result when $nl/p \ll 1$, so \eqref{eq:probOverlap} is fairly accurate.

Note that \eqref{eq:evenprob} and \eqref{eq:probOverlap} are identical.

An approximation of \eqref{eq:probOverlap} can be obtained from the work of \citet[p.~69]{Knuthvol2}.
He derived \eqref{eq:KnuthProbColl}, which is an upper bound on the average total number of
collisions~$c$ of $n$~balls thrown into $m$~urns.
Let us divide the period (of length $p$) of a generator into equal-sized
adjacent blocks of length $2l$, where the factor of 2 accounts for vulnerable zones.
Substituting the number of such blocks $m=p/2l$ into \eqref{eq:KnuthProbColl}
yields \eqref{eq:KnuthProbOver}, which is close to \eqref{eq:probOverlap}.
When $c$ is small, it is an upper bound on the probability of any overlaps occurring.
\begin{align}
   \label{eq:KnuthProbColl} & c < n^{2}/2m  \\
   \label{eq:KnuthProbOver} & \mathsf{P}(\text{overlap}) < ln^{2}/p .
\end{align}

\citet[p.~312]{FogVector}, \citet[p.~15]{LecuyerVector}, and \citet[``Remarks'' section]{VignaShootout}
have presented \eqref{eq:KnuthProbOver} in their work. Though \eqref{eq:KnuthProbOver} is more elegant,
\eqref{eq:probOverlap} is preferable because it is significantly more accurate when $n$ is small due to
running a job on a single multi-core computer.


\subsubsection{Generators with probabilistic periods}

The periods of the multiple cycles in a Romu generator are not known,
but they follow the probability distribution given in \eqref{eq:probCycleLen}.
To find the probability of overlap of its sequences,
we must convolve the distributions given in \eqref{eq:probCycleLen} and
\eqref{eq:probOverlap}. First, we list the definitions they need:

\begin{quote}
   $s$ = number of bits in the generator's state,\\
   $l$ = number of values in one sequence (stream),\\
   $n$ = number of such sequences,\\
   $r$ = $\log_{2}$(period of a randomly chosen cycle),\\
   $k$ = $\log_{2}$(period of a cycle).
\end{quote}

For a Romu generator, consider the cycle containing a randomly selected value $x$.
The probability that this cycle's period $r$ will be in the interval [$2^{k-1}$,$2^{k}$] is given by
\begin{equation} \label{eq:pinterval1}
   \mathsf{P}(r \in [k-1,k]) = 2^{k-s} - 2^{k-s-1} = 2^{k-s-1}.
\end{equation}

The convolution can then be expressed as follows:
\begin{align}
   \label{eq:conv1}
   \mathsf{P}(\mathrm{overlap}) & \approx \sum_{k=1}^{s} \mathsf{P}(r \in [k-1,k]) \cdot \mathsf{P}(\mathrm{overlap} \,|\, \mathrm{period~} p=2^{k-1/2}) \\
   \label{eq:conv2}
   & = \sum_{k=1}^{s} 2^{k-s-1} \frac{(n-1)nl}{2^{k-1/2}}  \\
   \label{eq:convApprox}
   & = \frac{(n-1)nls}{\sqrt{2} \cdot 2^s} .
\end{align}

The expression $2^{k-1/2}$ at the end of \eqref{eq:conv1} is the geometric mean of the interval of summation, [$2^{k-1},2^k$],
and was employed to improve accuracy. Despite this improvement,
the division of the range of $k$ into discrete intervals contributed some inaccuracy to \eqref{eq:convApprox}.
We can compute the probability more accurately using an integral.
We define $t = 2^{k-s}$ and note that $2^k = 2^{s}t$. With the appropriate substitutions,
the probability of any overlap occurring among $n$ sequences of $l$ values is

\begin{align}
   \mathsf{P}(\mathrm{overlap}) & \approx \int_{2^{-s}}^{1} \frac{(n-1)nl}{2^{s}t} \, \mathrm{d}t \notag \\
   \label{eq:convolved} & = \ln(2) \frac{(n-1)nls}{2^s} .
\end{align}

Equation~\eqref{eq:convolved} is used in the remainder of this paper.
Equation~\eqref{eq:convolved} may also be obtained by performing the summation in \eqref{eq:conv2}
using smaller steps, and then computing the limit as step-size is reduced to zero.
Equations \eqref{eq:convApprox} and \eqref{eq:convolved} differ by only 2\%, indicating that the geometric mean
is a close estimate. We must remember that, because \eqref{eq:probOverlap} is an upper bound, \eqref{eq:convolved}
is also an upper bound.

\begin{table}[ht!]
   \centering
   \begin{tabular}{ccccll}                                                                            \cline{5-6}
      \multicolumn{4}{c}{ \rule{0pt}{2.4ex} \phantom{x} }                     &  \multicolumn{2}{l}{Probability of overlap} \\ \hline
      \rule{0pt}{2.4ex}Row &           $s$  &   $l$    &    $n$               &  Known period            & Romu \\ \hline
      \rule{0pt}{2.7ex}1   &           256  & $2^{64}$ & $2^{40}$             &  \phantom{xxx}$2^{-112}$ & $2^{-104.5}$  \\
      2                    &           192  & $2^{58}$ & $2^{17}$             &  \phantom{xxx}$2^{-103}$ & $2^{-92.9}$   \\
      3                    &           128  & $2^{53}$ & $2^{14}$             &  \phantom{xxx}$2^{-47 }$ & $2^{-40.5}$   \\
      4                    & \phantom{0}96  & $2^{44}$ & $2^{8\phantom{0}}$   &  \phantom{xxx}$2^{-36 }$ & $2^{-29.9}$   \\
      5                    & \phantom{0}64  & $2^{44}$ & $2^{5\phantom{0}}$   &  \phantom{xxx}$2^{-10 }$ & $2^{-4.6}$    \\ \hline
   \end{tabular}
   \caption{\label{tab:convOverlap} Probabilities of sequence-overlap in generators.}
\end{table}

Table~\ref{tab:convOverlap} provides the probabilities of overlap for various generators and jobs.
The \emph{Known period} column contains probabilities for generators with known periods of about $2^s$, and is based on \eqref{eq:probOverlap}.
The probabilities in the \emph{Romu} column are based on \eqref{eq:convolved}.
For Romu generators:

\begin{itemize}
   \item 256 bits of state (row 1): This large state size far exceeds what could ever be needed.
   Absurdly large values of $l$ and $n$ result in an inconceivably low probability of overlap.

   \item 192 bits of state (row 2): The impossibly massive job in this row consists of
   131 thousand streams of $2^{58}$ values each.
   Consuming $2^{58}$ values at the unrealistically high rate of one per nanosecond would take 9 years.
   The resulting probability of overlap is approximately that of randomly selecting a specific snowflake
   among all snowflakes that have fallen in Earth's history.
   The human mind cannot conceive of such a low probability.
   Thus, in reality, streams will never overlap.

   \item 128 bits of state (row 3): The probability of overlap is under one in a trillion for a large job
   consisting of 16 thousand streams of $2^{53}$ values each.
   Optimistically assuming that each core consumes one value in a stream per nanosecond,
   this job would require 3.4 months of an entire computer-room.

   \item 96 bits of state (row 4): 96 state bits is barely adequate to supply a moderately large job
   with an overlap-probability of about one in a billion.

   \item 64 bits of state (row 5): Such a small state size is not sufficient for large applications,
   as a Romu generator cannot provide a large number of large streams with a low probability of overlap.
\end{itemize}

With at least 128 bits of state, we can safely ignore the possibility of overlap.

Some people might insist that a generator have exactly one period (of known length),
which eliminates nonlinear generators from consideration.
But with 192 or more bits of state, such a person will be unwisely losing the substantial benefit of improved performance
in order to avoid a will-never-happen risk of overlap (or a too-short cycle).
Randomly selecting one specific snowflake in Earth's history will never happen.
Losing benefits to avoid that nil risk is unwise.

The right two columns show that the convolution has raised the probabilities of overlap
for Romu generators (in relation to generators with known periods) by factors of $2^{5.5}$ to $2^{7.5}$.
For state sizes of at least 128 bits, the probabilities are so low that this boost is harmless.
But for state sizes of 64 or fewer bits, probabilities of overlap
are high enough that Romu generators should be avoided except for specialized applications.

\subsection{Equidistribution}

Some generators have the property of \emph{d-dimensional equidistribution}, which means that when
the generator is run over its entire period, all $d$-dimensional hypercubes will contain the same number
of points. A Romu generator is \emph{not} equidistributed because every cycle contains only a subset of the
$2^s$ possible states, causing hypercubes to contain an uneven number of points, as would occur with truly random numbers.
In fact, equidistribution can be regarded as a \emph{defect} in a generator because truly random numbers are
not equidistributed.


\citet[p. 203]{LecuyerEqui} claims the following benefits of equidistribution:
``Such generators have strong theoretical support and lend themselves to very fast
software implementations.'' Unfortunately, the strong theoretical support applies only to linear
generators, and not to nonlinear generators such as Romu. And the claim of ``very fast'' is
suspect because our testing shows that the nonequidistributed Romu generators
are faster than equidistributed generators with the same degree of randomness.
\citet{BrentEquidist} stated his conclusion clearly:

\begin{quote}
    When comparing modern long-period pseudo-random number generators, $(d,w)$-equidistribution is irrelevant,
    because it is neither necessary nor sufficient for a good generator.
\end{quote}


\section{Creating and Using Romu Generators} \label{sec:createUse}

\subsection{Determination of constants}

The Romu generators presented in this paper contain some constant scalars, specifically,
rotation-counts and multipliers. These could not be determined mathematically because
the nonlinear arithmetic in Romu generators is mathematically intractable, as mentioned above.
And they could not be determined empirically with the actual generators because doing so would
take centuries.

The solution was to employ the mini-generators described above. However, in some cases, mini-generators
with 40 and 48 bits of state were created when 32 bits was found to be too small and 64 bits was too large.
Using \pkg{PractRand}, each trial rotation-count was scored based on
the amount of data required to produce a failure. Generators containing
two rotations required evaluating many trial rotation-pairs.
Preference was given to high-scoring rotation-count(s) whose neighbors (with counts differing by 1 or 2)
also had high scores, indicating that the prospective rotation-count(s) was not on a narrow peak,
but was at the top of a gentle hill, providing assurance that the scaled-up count would also be at or near the peak.
Each prospective rotation-count(s) was also tested with multiple multipliers to ascertain
its sensitivity to multipliers.
The winning rotation-count(s) was then scaled up for the full-size generator.

The above procedure cannot determine multipliers because they cannot be scaled up. Instead,
multipliers were created randomly and then hand-modified using the following heuristics:

\begin{enumerate}
   \advance\parskip -4pt
   \item The 4 least significant bits (LSBs) were set to 1011 to (1) make the multiplier odd
         so that multiplications would be invertible, and (2) mix the LSBs of multiplicands well.
   \item The multiplier's Hamming weight (i.e., its number of 1-bits) was changed to be near
         $w/2$, where $w$ is the word-size of the generator's arithmetic (64 or 32).
   \item Any repeated patterns in the multiplier's bits were broken up.
   \item Long runs of 0- or 1-bits were avoided.
\end{enumerate}

Several prospective 64-bit multipliers were then tested using a RomuMono generator that was
modified to output 56 bits instead of the usual 32, making it poor enough that \pkg{PractRand} would
fail it in a few minutes. All of the multipliers performed similarly, even with different rotations. The
ratio of the best number of bytes processed to the worst was about 8.0, which is insignificant
at large magnitudes. The best multiplier was selected for use in all of the 64-bit Romu generators
presented in this paper.

\subsection{Seeding Romu generators}

Every Romu generator presented in this paper must be seeded such that its initial state
is not all zeros. Any nonzero seed will suffice, including one that is almost all
zeros. A seed with all zeros except for one 1-bit is acceptable. Due to its multiplication and
rotation(s), a Romu generator quickly escapes from an initial mostly-zeros state.
Also, streams that are seeded with a counter (1, 2, 3, \ldots) will not exhibit correlation.

\subsection{Rotation operation}

Because Romu generators perform rotations, we define the following rotate-left macro in \proglang{C/C++}:
\begin{Code}
#define ROTL(d,lrot)  ((d<<(lrot)) | (d>>(8*sizeof(d)-(lrot))))
\end{Code}

Modern compilers such as gcc and Visual Studio\texttrademark{} recognize the expression in this macro as a
left-rotation, and emit one instruction for it. Also, the \code{sizeof} operator in this macro enables
it to properly rotate operands of any size.
All \proglang{C}-code examples provided in this paper use this ROTL macro.
The following is the equivalent template-function in \proglang{C++}:
\begin{Code}
template <class uDataT>
inline uDataT rotl (uDataT d, unsigned lrot) {
   return (d<<lrot) | (d>>(8*sizeof(d)-lrot));
}
\end{Code}

\section{Romu Generators using 64-bit Arithmetic}

\subsection{Introduction to 64-bit generators}

The Romu generators employing 64-bit arithmetic are presented below.
Their traits are summarized in Table~\ref{tab:summary64},
along with the traits of some of their competitors.

\begin{table}[ht!]
   \centering
   \begin{tabular}{ l @{} cclccccc }  \hline
      \rule{0pt}{2.4ex}            & Bits of  & Output & \multicolumn{1}{c}{Arith.} & Proc. & Register &  Tested   &    Capacity                       \\
      \multicolumn{1}{c}{Generator}& state    & latency& \multicolumn{1}{c}{ops}    & cycles& pressure &  bytes    &    (bytes)                        \\  \hline
      \rule{0pt}{2.6ex}RomuQuad    & 256      &    0   &             \phantom{i11}6 &   3   &    8     & $2^{48}$  &  $\ge 2^{90}$ (est.)              \\
      RomuTrio                     & 192      &    0   &             \phantom{i11}5 &   3   &    6     & $2^{48}$  &  \phantom{$\ge$} $2^{75}$ (est.)  \\
      RomuDuo                      & 128      &    0   &             \phantom{i11}5 &   3   &    5     & $2^{48}$  &  \phantom{$\ge$} $2^{61}$ (est.)  \\
      RomuDuoJr                    & 128      &    0   &             \phantom{i11}3 &   3   &    4     & $2^{48}$  &  \phantom{$\ge$} $2^{51}$ (est.)  \\
      xoshiro256++                 & 256      &    3   &             \phantom{i1}10 &   3   &   7-8    & $2^{45}$  &                                   \\
      xoroshiro128++               & 128      &    3   &             \phantom{i11}9 &   3   &    5     & $2^{45}$  &                                   \\
      pcg-xsh-rs (128\textrightarrow 64) & 128&    7   &             \phantom{i1}17 &   9   &   10     & $2^{42}$  &  \phantom{$\ge$} $2^{42}$ \phantom{(est.)} \\ \hline
   \end{tabular}
   \caption{\label{tab:summary64} Summary of 64-bit generators.}
\end{table}

In Table~\ref{tab:summary64}:

\emph{Bits of state} is 64 multiplied by the number of state variables in the generator (2, 3, or 4).

\emph{Output latency} is an important concept. It is the delay (in clock cycles) the generator
imposes on the application when the generator is inlined in a superscalar processor. In a well
designed generator, the output value becomes available before the generator's processing is done,
allowing the application to execute in parallel with the generator. The 64-bit
Romu generators presented in this paper have no output latency, and thus cause no delay in the application.
From the application's point of view, these generators consume no time.  

As an aside, any generator that computes its output-value (and thus has a nonzero output latency) can be
modified to have no output latency by computing and saving the next output-value in a variable to be returned
in the next call or invocation. But this trick comes with a cost: It boosts register pressure. The additional
memory-accesses due to a spill will reduce or erase the time-savings.

\emph{Arith.\ ops} is the number of arithmetic operations performed by the generator, excluding
data-movement. Note that the Romu generators have around half or fewer operations than their (linear)
competitors, revealing the power of a rotation in conjunction with a multiplication. The pcg-xsh-rs
generator has a large number of operations because each of its 128-bit operations creates multiple
64-bit operations; these were counted in the compiler's assembler-language output. \emph{Arith.\ ops}
is also a rough indication of how fast the generator will run on a nonsuperscalar processor.

\emph{Proc.\ cycles} is the number of clock cycles of processing time needed by the generator when inlined,
assuming that its variables are held in registers. When this value is greater than \emph{Output latency},
ILP is causing the application and generator to execute concurrently after the random number has been output.
When the generator is not inlined, it is being called as a function,
so the time spent reading/writing variables and the call/return overhead must
be added to this processing time.

\emph{Register pressure} is the number of registers devoted to the generator when inlined in a
loop. It is important to count registers in a loop and not in a function, as a function usually
uses fewer registers because writes to memory also free registers. These numbers were
determined by compiling with g++ version 4.8.4 with the \code{-S -O3} options,
and examining the resultant assembler-language file.

\emph{Tested bytes} is the number of bytes (not number of values) supplied to \pkg{PractRand}
in a long test. Each generator passed with no signs of stress.

\emph{Capacity} is the maximum number of bytes each generator can supply before statistical
anomalies become apparent in \pkg{PractRand}. When this
value exceeds \emph{Tested bytes}, it is an estimate (marked ``(est.)'' in the table) based on testing a scaled-down version of
the generator as described above. A value of the form ``$\ge 2^n$'' indicates that the scaled-down
generator passed its test without stress or failure. These estimates are conservative, so the
true maxima should be higher.

\subsection{RomuQuad generator}

The state of RomuQuad comprises four 64-bit variables (256 bits). Its processing time is 3 cycles, and it
has no output latency, allowing an application to execute as soon as possible. This is the best
64-bit Romu generator presented in this paper, but it also has the highest register pressure, which is 8.
RomuTrio is recommended instead due to its lower register pressure of 6.
But RomuQuad can be employed for massive jobs run by users who choose to be extremely cautious
about the probability of overlap.

\begin{Code}
uint64_t wState, xState, yState, zState;  // set to nonzero seed

uint64_t romuQuad_random () {
   uint64_t wp = wState, xp = xState, yp = yState, zp = zState;
   wState = 15241094284759029579u * zp; // a-mult
   xState = zp + ROTL(wp,52);           // b-rotl, c-add
   yState = yp - xp;                    // d-sub
   zState = yp + wp;                    // e-add
   zState = ROTL(zState,19);            // f-rotl
   return xp;
}
\end{Code}

The ILP table (below) for RomuQuad reveals a subtle reason this generator will slightly slow down the application.
All four slots are occupied in cycle 1, causing the generator and the application to compete for one of those slots,
delaying one of them by a cycle.
On average, the application will be delayed by half a clock cycle, which is insignificant.

\begin{center}
   \begin{tabular}{cccccc}  \hline
      \rule{0pt}{2.4ex}Cycle &  Slot 1          & Slot 2           & Slot 3           &  Slot 4          &  Multiplier  \\  \hline
      \rule{0pt}{2.5ex}1     & b-rotl           & d-sub            & e-add            &  a-mult          &      *       \\
                       2     & c-add            & f-rotl           &                  &                  &      *       \\
                       3     & \phantom{xxxxxx} & \phantom{xxxxxx} & \phantom{xxxxxx} & \phantom{xxxxxx} &      *       \\  \hline
   \end{tabular}  
\end{center}

One might regard RomuQuad as too large, but its half-size version (RomuQuad32)
presented later requires all four 32-bit state variables in order to reach 128 bits of state.

\begin{table}[ht!]
   \centering
   \begin{tabular}{cccl}  \hline
      \rule{0pt}{2.4ex} Output variable & Input var. 1 & Input var. 2 & Transform    \\  \hline
                      \code{wState}     & \code{zp}    &              & multiply     \\
                      \code{xState}     & \code{wp}    & \code{zp}    & rotate, add  \\
                      \code{yState}     & \code{xp}    & \code{yp}    & subtract     \\
                      \code{zState}     & \code{yp}    & \code{wp}    & add, rotate  \\  \hline
   \end{tabular}
   \caption{\label{tab:dataflows} Dataflows in RomuQuad.}
\end{table}


The dataflows from input variables to output variables are instructive.
Column \emph{Input var.~1} of Table~\ref{tab:dataflows} shows that
each output variable is computed from the prior input variable.
For example, \code{yState} is computed from \code{xp}.
Drawing from the prior input variable ensures that data flows circularly among the four state-variables.
If those were the only dataflows, then after every four calls, no variable would have affected any
other, so in essence there would be four independent 64-bit generators operating in tandem,
creating poor randomness. Therefore, every output variable (except \code{wState)} is also computed
from a second input variable (in column \emph{Input var.~2}), causing the variables to mix with each other.
Except for \code{wState}, every output variable is computed from two input variables.
Furthermore, every input variable (except \code{yp}) supplies two output variables. This two-out/two-in design mixes
the state variables well. A rotation contributes to every other output variable (\code{xState} and \code{zState}),
creating an alternation between rotations and additions, improving the mixing of bits.

An alternate design is to have every output variable use the prior two input variables, except
for the product in \code{wState}. For example, \code{yState} would be computed from \code{wp} and \code{xp}.
Tests on a 32-bit generator
consisting of four 8-bit state variables indicated that this approach matches the randomness
of RomuQuad. This design was not investigated further as it improves neither performance nor
randomness, and it was feared that the repeating prior-two-into-one pattern might produce subtle
patterns in output values.

\subsection{RomuTrio generator}

RomuTrio has 192 $(3 \! \times \! 64)$ bits of state, and as seen in Table \ref{tab:convOverlap},
such a generator can create many huge sequences with essentially
no chance of overlap (one snowflake in Earth's history). Furthermore, RomuTrio easily passes
all statistical tests. RomuQuad is fast, and RomuTrio is faster due to its lower register
pressure and sparser ILP table. For these reasons, this generator is recommended most highly.

\begin{Code}
uint64_t xState, yState, zState;  // set to nonzero seed

uint64_t romuTrio_random () {
   uint64_t xp = xState, yp = yState, zp = zState;
   xState = 15241094284759029579u * zp;
   yState = yp - xp;  yState = ROTL(yState,12);
   zState = zp - yp;  zState = ROTL(zState,44);
   return xp;
}
\end{Code}

The ILP Table (below) for RomuTrio reveals that it makes excellent use of ILP. Note that slot 4 is empty
in all three cycles, allowing the application to execute in parallel with this generator,
maximizing its speed.

\begin{center}
   \begin{tabular}{cccccc}  \hline
      \rule{0pt}{2.4ex}Cycle & Slot 1               & Slot 2               & Slot 3               & Slot 4               &  Multiplier  \\  \hline
      \rule{0pt}{2.5ex}1     & subtract x           & subtract y           & multiply             &                      &      *       \\
                       2     & rotate y             & rotate z             &                      &                      &      *       \\
                       3     & \phantom{subtract x} & \phantom{subtract x} & \phantom{subtract x} & \phantom{subtract x} &      *       \\  \hline
   \end{tabular}  
\end{center}

As seen in Table~\ref{tab:summary64}, the processing time of xoshiro256++ (by Blackman and Vigna)
matches that of RomuQuad and RomuTrio, but its output latency is 3 cycles, a noticeable speed loss in time-critical applications.
The ILP table (below) for xoshiro256++ reveals the genius behind it, as all four slots have been
thoughtfully filled. But this full ILP table could cause some speed loss if the application inserts
an instruction or two in these slots, pushing the generator's instruction(s) into additional
clock cycle(s). Overall, xoshiro256++ is slower than RomuTrio due to its 3-cycle output latency,
fragile ILP-table, and higher register pressure.

\begin{center}
   \begin{tabular}{cccccc}  \hline
      \rule{0pt}{2.4ex}Cycle & Slot 1               & Slot 2               & Slot 3               & Slot 4               &  Multiplier  \\  \hline
      \rule{0pt}{2.5ex}1     & add                  & shift                & xor                  & xor                  &              \\
                       2     & rotate               & xor                  & xor                  & xor                  &              \\
                       3     & add                  & rotate               & \phantom{rotate}     & \phantom{rotate}     &              \\  \hline
   \end{tabular}  
\end{center}

The xoshiro256++ generator has the advantage of offering a jump-ahead function, which is
supplied with the generator's source-code. No Romu generator can have this feature due to
its intractable nonlinear arithmetic. But does this feature help an application? Jump-ahead is used to
create nonoverlapping streams. The 192 bits of state in RomuTrio ensure that its streams will
not overlap, eliminating the need for jump-ahead.

\subsection{RomuDuo generator}

RomuDuo will be faster than RomuTrio when its lower register pressure causes fewer memory-spills.
Its state size of 128 bits is large enough to support many streams with essentially no chance of overlap.
The estimated capacity of RomuDuo is $2^{58}$ values ($2^{61}$ bytes), which is significantly lower than that of RomuTrio or RomuQuad.
But this lower capacity is still larger than any job, and it would take 9 years to consume it at the rate of one per nanosecond.
This estimate of capacity should be too low because it is conservative. But because it is an estimate and not a measurement,
a prudent worker will maintain some margin between job/stream-size and capacity for large jobs requiring a high degree of randomness.

\begin{Code}
uint64_t xState, yState;  // set to nonzero seed

uint64_t romuDuo_random () {
   uint64_t xp = xState;
   xState = 15241094284759029579u * yState;
   yState = ROTL(yState,36) + ROTL(yState,15) - xp;
   return xp;
}
\end{Code}

The ILP table (below) for RomuDuo is sparse, leaving plenty of slots for the application to use, which can only improve performance.

\begin{center}
   \begin{tabular}{cccccc}  \hline
      \rule{0pt}{2.4ex}Cycle & Slot 1               & Slot 2               & Slot 3               & Slot 4               &  Multiplier  \\  \hline
      \rule{0pt}{2.5ex}1     & rotate               & rotate               & multiply             &                      &      *       \\
                       2     & add                  &                      &                      &                      &      *       \\
                       3     & subtract             & \phantom{subtract}   & \phantom{subtract}   & \phantom{subtract}   &      *       \\  \hline
   \end{tabular}  
\end{center}

\subsection{RomuDuoJr generator}

RomuDuoJr is a simplification of RomuDuo, removing a rotation and addition.
Hence, we appended Jr to its name, which stands for ``Junior''.
The reduced number of operations and register pressure make it the fastest
generator presented in this paper using 64-bit arithmetic.
RomuDuoJr is suitable for most applications, and it should be preferred when speed is paramount.
However, a large job can exceed its reduced capacity, so one must be careful.

\begin{Code}
uint64_t xState, yState;  // set to nonzero seed

uint64_t romuDuoJr_random () {
   uint64_t xp = xState;
   xState = 15241094284759029579u * yState;
   yState = yState - xp;  yState = ROTL(yState,27);
   return xp;
}
\end{Code}

The ILP table (below) for RomuDuoJr is even sparser than that of RomuDuo. In fact, as pointed out
in Section~\ref{sec:capacity}, this generator consists of only three arithmetic instructions.

\begin{center}
   \begin{tabular}{cccccc}  \hline
      \rule{0pt}{2.4ex}Cycle & Slot 1               & Slot 2               & Slot 3               & Slot 4               &  Multiplier  \\  \hline
      \rule{0pt}{2.5ex}1     & subtract             & multiply             &                      &                      &      *       \\
                       2     & rotate               &                      &                      &                      &      *       \\
                       3     &                      & \phantom{subtract}   & \phantom{subtract}   & \phantom{subtract}   &      *       \\  \hline
   \end{tabular}  
\end{center}

We can compare O'Neill's PCG \citep{pcg} to RomuDuo/RomuDuoJr. The PCG is an LCG with a randomized
scrambler added as a post-processing step. The pcg-xsh-rs variant uses 128-bit-wide arithmetic,
and returns 64 bits of output.
Its state-size is 128 bits, matching RomuDuo/RomuDuoJr.
Its capacity is $2^{42}$ bytes of output.
But its 128-bit-wide arithmetic makes it much slower than all Romu generators,
as Table~\ref{tab:summary64} shows.
Furthermore, its large register pressure of 10 will cause memory-spills, slowing the application even more.
Like the Blackman-Vigna generators, the PCG offers a jump-ahead function. And as mentioned above,
jump-ahead is unnecessary for generators with 128 or more bits of state due to the
extremely low probabilities of sequence-overlap.

\section{Romu Generators using 32-bit Arithmetic}

32-bit generators are needed when running on specialized hardware lacking 64-bit arithmetic,
such as the Intel Nios II \textregistered{} soft core and DSP chips such as the Texas Instruments C66x
family. Their small word-size of 32-bits makes it difficult for their generators to have a
large number of state-bits without resorting to an internal array, which would reduce performance due
to reading/writing memory and managing indices or pointers.

RomuQuad32 is suitable for large jobs, and RomuTrio32 and RomuMono32 are 
suitable for medium-to-large and small jobs, respectively.
Among these choices, RomuTrio32 is the fastest, so we recommend
preferring it for jobs that will not exceed its capacity.

%

\begin{table}[ht!]
   \centering
   \begin{tabular}{l @{} cclccc @{ } c}  \hline
      \rule{0pt}{2.4ex}                                  & Bits of          & Output  &          Arith. & Proc.  &  Register  &  Tested   &    Capacity  \\
      \multicolumn{1}{c}{Generator}                      & state            & latency & \phantom{x}ops  & cycles &  pressure  &  bytes    &    (bytes)   \\  \hline
      \rule{0pt}{2.6ex}RomuQuad32                        &           128    &    0    & \phantom{00}6   &   3    &     7      & $2^{45}$  & $\ge 2^{62}$ (est.)\\
                       RomuTrio32                        & \phantom{0}96    &    0    & \phantom{00}5   &   3    &     5      & $2^{45}$  & $2^{53}$ (est.)\\
                       RomuMono32 (32\textrightarrow 16) & \phantom{0}32    &    1    & \phantom{00}2   &   4    &     2      & $2^{27}$  & $2^{27}$     \\
                       xoshiro128$**$                    &           128    &    5    & \phantom{0}10   &   5    &     6-7    & $2^{45}$  &              \\
                       pcg-xsh-rr (32\textrightarrow 16) & \phantom{0}32    &    4    & \phantom{00}7   &   4    &     3      & $2^{27}$  & $2^{27}$     \\
                       LCG (32\textrightarrow upper 16)  & \phantom{0}32    &    0    & \phantom{00}2   &   4    &     2      & $2^{17}$  & $2^{17}$     \\  \hline
   \end{tabular}
   \caption{\label{tab:summary32} Summary of 32-bit generators.}
\end{table}

\subsection{RomuQuad32 generator}

This 128-bit generator is suitable for large jobs, is very fast (with no output latency),
and has excellent statistical quality.
It is the only Romu generator using 32-bit arithmetic this paper recommends for general purpose use.
However, as mentioned above, RomuTrio32 is faster due to its lower register pressure, so we recommend
employing it when feasible.

Earlier, it was stated that a register pressure of 7 or more is regarded as excessive. But a state
size of 128 bits requires that four 32-bit variables be kept in registers, and after adding
the necessary temporary registers, high register pressure became unavoidable in RomuQuad32.
The register pressure of RomuQuad32 is lower than that of RomuQuad because
a quirk in the Intel/AMD instruction set forces a register to be dedicated to holding a 64-bit
multiplier, an aberration not present in its 32-bit instructions.

\begin{Code}
uint32_t wState, xState, yState, zState;  // set to nonzero seed

uint32_t romuQuad32_random () {
   uint32_t wp = wState, xp = xState, yp = yState, zp = zState;
   wState = 3323815723u * zp;  // a-mult
   xState = zp + ROTL(wp,26);  // b-rotl, c-add
   yState = yp - xp;           // d-sub
   zState = yp + wp;           // e-add
   zState = ROTL(zState,9);    // f-rotl
   return xp;
}
\end{Code}

The ILP table (below) is identical to that of RomuQuad because this is a half-size version of RomuQuad.

\begin{center}
   \begin{tabular}{cccccc}  \hline
      \rule{0pt}{2.4ex}Cycle &  Slot 1          & Slot 2           & Slot 3           &  Slot 4          &  Multiplier  \\  \hline
      \rule{0pt}{2.5ex}1     & b-rotl           & d-sub            & e-add            &  a-mult          &      *       \\
                       2     & c-add            & f-rotl           &                  &                  &      *       \\
                       3     & \phantom{xxxxxx} & \phantom{xxxxxx} & \phantom{xxxxxx} & \phantom{xxxxxx} &      *       \\  \hline
   \end{tabular}  
\end{center}

Blackman and Vigna's xoshiro128$**$ generator competes against RomuQuad32. Both have 128
bits of state using 32-bit arithmetic with about the same register pressure, producing top
quality values. But Table~\ref{tab:summary32}  shows that RomuQuad32 is 5 cycles faster. Though xoshiro128$**$ is slower,
it has the advantage of performing \emph{no} multiplications (despite the stars in its name)
because it multiplies by the constants 5 and 9 which the compiler translates into \code{lea} instructions
or shifts-and-adds. If multiplication is slow or absent in the processor,
xoshiro128$**$ is probably the best choice. Otherwise, the higher speed of RomuQuad32 makes it preferable.

\subsection{RomuTrio32 generator}

With 96 bits of state, this generator is not suitable for the largest jobs, as Table~\ref{tab:convOverlap} reveals.
But it can supply the needs of most applications. Because its register pressure is lower than RomuQuad32,
there will be fewer spills, causing an application to run faster.

\begin{Code}
uint32_t xState, yState, zState;  // set to nonzero seed

uint32_t romuTrio32_random () {
   uint32_t xp = xState, yp = yState, zp = zState;
   xState = 3323815723u * zp;
   yState = yp - xp;  yState = ROTL(yState,6);
   zState = zp - yp;  zState = ROTL(zState,22);
   return xp;
}
\end{Code}

Because this generator is a half-size version of RomuTrio, its ILP table (below) is identical.
The empty slot 4 allows the application to run at maximum possible speed because it will
never wait on the generator.

\begin{center}
   \begin{tabular}{cccccc}  \hline
      \rule{0pt}{2.4ex}Cycle & Slot 1               & Slot 2               & Slot 3               & Slot 4               &  Multiplier  \\  \hline
      \rule{0pt}{2.5ex}1     & subtract x           & subtract y           & multiply             &                      &      *       \\
                       2     & rotate y             & rotate z             &                      &                      &      *       \\
                       3     & \phantom{subtract x} & \phantom{subtract x} & \phantom{subtract x} & \phantom{subtract x} &      *       \\  \hline
   \end{tabular}  
\end{center}

\subsection{RomuMono32 generator}

We first published this elegant generator in \citet{OvertonDDJ}, but named it ``CMR'',
which stands for ``constant, multiply, rotate'', because it multiplies by a constant
and rotates the result.
RomuMono32 is tiny, consisting of only those two arithmetic operations plus one more to extract the returned result.
Because RomuMono32 has only 32 bits of state, the period of its longest cycle can be traversed in a few seconds,
allowing one to search for multipliers and rotation-counts that yield cycles with periods
very close to $2^{32}$. If $p$ is the period resulting from such a multiplier-rotation pair, let
$d = 2^{32} - p$, which is the amount the period falls short of the ideal of $2^{32}$. For several months,
all cores of several computers randomly searched for multiplier-rotation pairs that yield
small values of $d$.

If $d$ is small (i.e., under about 10000), and if one were to sort all numbers in the longest cycle,
there would be large blocks of numbers that are consecutive on the number line.
These can be used to seed the generator, guaranteeing that every seed from the block will be in the longest cycle.
When seeded in this manner, the generator's period is known and is not probabilistic.
The length of the largest such \emph{seed-block}, as we term them, determines the number
of bits of seed the generator can accept. The mean length of a seed-block is $2^{32}/d$. For example,
if $d$ is 47, seed-blocks will average $2^{26.4}$ values. The largest seed-block will be larger due
to random variation in their sizes, so such a generator can accept more than 26 bits of seed,
as seen in function \texttt{romuMono32\_init} below.

\begin{Code}
uint32_t state;

void romuMono32_init (uint32_t seed) {
   state = (seed & 0x1fffffffu) + 1156979152u;  // Accepts 29 seed-bits.
}

uint16_t romuMono32_random () {
   uint16_t result = state >> 16;
   state *= 3611795771u;  state = ROTL(state,12);
   return result;
}
\end{Code}

The $d$-value of this generator is 47, so its period is $2^{32}\!-\!47$. The unusual method of
seeding in \texttt{romuMono32\_init} adds the 29 LSBs of the seed to the beginning of the largest
seed-block for the given multiplier and rotation.

Appendix~\ref{app:mrtable} lists all multiplier-rotation pairs having $d$-values under 2000
that the long search found. Rotations up to only 16 are included because testing revealed
that randomness suddenly drops with larger rotations. However, despite its reduced randomness,
the second row in the table with a rotation-count over 16 was included due to its extraordinarily low $d$-value.

The period of RomuMono32 is about $2^{32}$, which is small by today's standards, so it should be
used only for small applications. This small period creates the temptation to combine two or
three such generators, perhaps by mixing their outputs with an add or xor, yielding periods of
$2^{64}$ or $2^{96}$. This experiment was done, and it was found that two RomuMono32 generators in combination
can output 32 bits each (instead of the usual 16), creating 32-bit combined outputs which
will pass \pkg{BigCrush} and \pkg{PractRand} at $2^{45}$ bytes. When combining these generators, one must ensure
that their periods are coprime in order to achieve the longest possible combined period.
It is also prudent to choose rotations that differ by at least~2.

But such a combination generator has a problem: There is likely to be detectable correlation
between portions of two sequences with lengths approaching $2^{31}$. And for lengths exceeding
$2^{31}$, such correlated portions will surely exist because a base generator must have contributed
the same subsequence to both output sequences, causing detectable correlation between them. This
is the reason no combination RomuMono32 generator was presented in this paper.

Though this tiny generator employs only two arithmetic operations to update its state,
it matches the statistical quality of the 32-bit pcg-xsh-rr variant of the PCG.
The capacity of both is $2^{26}$ 16-bit values. But the PCG requires 7 arithmetic
operations to accomplish this feat, versus 3 for RomuMono32. The output latency of the PCG is 4
cycles versus 1 for RomuMono32, making it much faster. The purpose of saying this is not to denigrate
the PCG, as it is a good generator; the purpose is to emphasize that a mixture of linear and
nonlinear arithmetic can produce a good generator with significantly fewer arithmetic operations.

Table~\ref{tab:summary32} reveals this advantage in its comparison of RomuMono32 with an LCG that
returns its high-order 16 bits. Both consist of 3 arithmetic instructions. Both execute in the
same amount of time. Both have 32 bits of state. But the LCG can output only $2^{16}$ values
before troubles become apparent, versus $2^{26}$ values for RomuMono32.

That $2^{26}$ is large enough to be useful. When a small job running in a 32-bit processor
needs a tiny generator, RomuMono32 is a good choice.


\section{Conclusion} \label{sec:summary}

Members of the Romu family offer the following advantages over other good generators:
\begin{itemize}
   \advance\parskip -3pt
   \item They are faster. Except for RomuMono32, they consume only 3 clock cycles due to good ILP,
         and when inlined, they cause \emph{no} delay in applications, maximizing their speed.
   \item They are more random for a given number of arithmetic operations.
   \item Their approximate capacities are known, so they do not impose unknown risks to jobs.
\end{itemize}

\begin{table}[ht!]
   \centering
   \begin{tabular}{lccc}  \hline
\rule{0pt}{2.4ex}\phantom{\checkmark}Generator  & Capacity (bytes) & Bits of state  &  Arithmetic \\  \hline
\rule{0pt}{2.4ex}\phantom{\checkmark}RomuQuad   & $2^{90}$         &       256      &     64      \\          
                   \checkmark RomuTrio          & $2^{75}$         &       192      &     64      \\
          \phantom{\checkmark}RomuDuo           & $2^{61}$         &       128      &     64      \\          
          \phantom{\checkmark}RomuDuoJr         & $2^{51}$         &       128      &     64      \\  \hline
\rule{0pt}{2.4ex}\phantom{\checkmark}RomuQuad32 & $2^{62}$         &       128      &     32      \\          
                 \phantom{\checkmark}RomuTrio32 & $2^{53}$         & \phantom{0}96  &     32      \\
          \phantom{\checkmark}RomuMono32        & $2^{27}$         & \phantom{0}32  &     32      \\  \hline
   \end{tabular}
   \caption{\label{tab:recgens} Summary of Romu generators.}
\end{table}

Table~\ref{tab:recgens} summarizes the Romu generators presented in this paper.
RomuTrio has a check mark because it is most highly recommended for general purpose use,
as it offers both high speed and infinitesimal probabilities of too-short cycles or stream-overlap.
The second recommendation is RomuQuad, but it has higher register pressure and its improved probabilities
(compared to RomuTrio) will make no difference in practice. RomuDuo and RomuDuoJr might be even faster
than RomuTrio due to lower register pressure, and their probabilities of overlap are low,
but the largest jobs might approach the capacity of RomuDuo, and will exceed that of RomuDuoJr.
For 32-bit processors, RomuQuad32 is intended for general purpose work, RomuTrio32 can supply most jobs,
and RomuMono32 is suitable only for small jobs.

It appears that theoreticians avoid nonlinear generators because their periods and structural traits cannot
be determined mathematically. However, their periods are probabilistic, and the convolution
presented in this paper shows that, with a state size of at least 128 bits, the probability of
a too-short cycle or stream-overlap in a large job will be vanishingly low,
causing any user to confidently say, ``It will never happen.''

Arithmetic that combines the linear operation of multiplication with the nonlinear
operation of rotation is inherently more random than employing only linear operations. For
example, using only three arithmetic instructions, no 64-bit linear generator can match the
randomness of RomuDuoJr. And using only two arithmetic instructions, no linear generator can
match the randomness of RomuMono. Because they perform fewer operations, Romu generators
are faster than the equivalent linear generators. 

We conclude that the invertible nonlinear generators in the Romu family are superior to
linear generators, and we hope researchers can improve upon those presented in this paper.

\section*{Acknowledgements}

We would like to thank Chris Doty-Humphrey for creating the \pkg{PractRand} suite of tests for
generators. It is the only test suite that provides a systematic quantitative measurement of
the quality of a generator. It performed all quality measurements needed by this paper.


\bibliography{romupaper}


\newpage
\begin{appendix}
\section{Good Multiplier-Rotation Pairs for RomuMono32} \label{app:mrtable}

In the table below, the two \emph{RM} columns apply when the rotation is done before the multiplication,
and the \emph{MR} columns apply when that order is reversed. The randomness of the generator is the
same for both orders. The two \emph{base} columns are the lowest values in the largest seed-blocks,
and the \emph{bits} columns are the corresponding numbers of bits of seed those blocks can accommodate.

\begin{tabular}{ccrrcrc}  \hline
\rule{0pt}{2.4ex}Multiplier & Lrot & $d$-value & RM-base\phantom{0} & RM-bits & MR-base\phantom{0} & MR-bits  \\ \hline
\rule{0pt}{2.4ex}2540121707  &  14  &     2\phantom{0}  &  437125826   &  31  &           1  &  31  \\
                 3731015275  &  18  &     3\phantom{0}  &          1   &  31  &  1564370705  &  30  \\
                 2336447867  &  16  &    43\phantom{0}  &  3779345575  &  28  &  2076771216  &  29  \\
                 3611795771  &  12  &    47\phantom{0}  &  342645537   &  28  &  1156979152  &  29  \\
                 3952805931  &  13  &   157\phantom{0}  &  1159051389  &  26  &  2515406761  &  26  \\
                 3276993211  &  14  &   184\phantom{0}  &  226123367   &  27  &  1335894518  &  27  \\
                 4084487243  &  12  &   347\phantom{0}  &  1339372056  &  26  &  2141099809  &  25  \\
                 4127380763  &  12  &   397\phantom{0}  &  606354474   &  26  &  1315333761  &  25  \\
                 3563976171  &  16  &   420\phantom{0}  &  2941035005  &  25  &  1377002680  &  25  \\
                 3651999659  &  14  &   429\phantom{0}  &  3209498982  &  25  &  2227458680  &  26  \\
                 3365008619  &  12  &   476\phantom{0}  &  874642096   &  25  &  2055179747  &  26  \\
                 3953463755  &  16  &   514\phantom{0}  &  4174029353  &  26  &  2817505743  &  25  \\
                 3989591211  &  13  &   592\phantom{0}  &  3721670923  &  25  &  3643450503  &  25  \\
                 3332453915  &  14  &   650\phantom{0}  &  3818020724  &  25  &  1456733700  &  25  \\
                 3586487947  &  16  &   681\phantom{0}  &  3925643197  &  25  &  3307401597  &  25  \\
                 4272641883  &  16  &   709\phantom{0}  &  3002109605  &  25  &  2625009597  &  25  \\
                 3525693099  &  14  &   758\phantom{0}  &  144357697   &  25  &  3440120341  &  25  \\
                 3690361499  &  14  &   768\phantom{0}  &  2120353896  &  24  &  2598302657  &  25  \\
                 1698147467  &  13  &   775\phantom{0}  &  1398136959  &  25  &  2483006683  &  25  \\
                 3319523819  &  15  &   838\phantom{0}  &  3085144041  &  25  &  2129925777  &  24 \\
                 3256525067  &  14  &   840\phantom{0}  &  1109021212  &  24  &  1180712204  &  25  \\
                 3552236683  &  13  &   890\phantom{0}  &  1495506820  &  25  &  2898009531  &  25 \\
                 3839154475  &  15  &   913\phantom{0}  &  1001325849  &  25  &  1262306147  &  24  \\
                 3636548587  &  13  &   920\phantom{0}  &  3509505418  &  25  &  3475628570  &  25 \\
                 4193921835  &  14  &   937\phantom{0}  &  1699403590  &  25  &  1473717306  &  24  \\
                 4074675915  &  15  &  1051\phantom{0}  &  1274742985  &  24  &  1207307984  &  25  \\
                 3589580459  &  16  &  1070\phantom{0}  &  2080133737  &  25  &  3604530546  &  24  \\
                 3937305835  &  15  &  1077\phantom{0}  &  1503300770  &  24  &  240484477   &  24  \\
                 3291786395  &  15  &  1121\phantom{0}  &  2583623346  &  24  &  652306397   &  24  \\
                 4070215643  &  15  &  1171\phantom{0}  &  4077664074  &  24  &  2173685177  &  24 \\
                 3603758123  &  12  &  1144\phantom{0}  &  3012437991  &  24  &  2526760669  &  25  \\
                 3475128667  &  14  &  1188\phantom{0}  &  2293176165  &  24  &  3061649031  &  24  \\
                 3852709339  &  15  &  1256\phantom{0}  &  2753678026  &  24  &  2610555647  &  24  \\
                 3316538187  &  13  &  1271\phantom{0}  &  2052216962  &  24  &  1133599252  &  24  \\
                 2759225787  &  12  &  1296\phantom{0}  &  566123425   &  24  &  3813660306  &  24  \\
                 3939093339  &  12  &  1336\phantom{0}  &  3269117755  &  24  &  2086898213  &  24  \\
                 3953739083  &  16  &  1338\phantom{0}  &  534924345   &  24  &  3580323618  &  24  \\
\end{tabular}

\begin{tabular}{ccrrcrc}  \hline
\rule{0pt}{2.4ex}Multiplier & Lrot & $d$-value & RM-base\phantom{0} & RM-bits & MR-base\phantom{0} & MR-bits  \\ \hline
\rule{0pt}{2.4ex}2662206315  &  15  &  1370\phantom{0}  &  3876606173  &  24  &  2061627875  &  24  \\
                 1422968075  &  16  &  1377\phantom{0}  &  332616298   &  24  &  3202323436  &  24  \\
                 4219099339  &  15  &  1379\phantom{0}  &  2712314424  &  24  &  3285589456  &  24  \\
                 3715414331  &  13  &  1388\phantom{0}  &  111908111   &  24  &  4000728640  &  24  \\
                 3662642315  &  14  &  1406\phantom{0}  &  781158169   &  25  &  2775093201  &  24  \\
                 3240747339  &  15  &  1443\phantom{0}  &  2061641557  &  24  &  109294508   &  24  \\
                 3505407659  &  16  &  1507\phantom{0}  &  2349523361  &  24  &  327552291   &  24  \\
                 3603832939  &  14  &  1535\phantom{0}  &  2430830656  &  24  &  883175065   &  24  \\
                 2334149515  &  15  &  1560\phantom{0}  &  2225828117  &  24  &  3989107689  &  24  \\
                 3804926571  &  12  &  1563\phantom{0}  &  3116634076  &  24  &  3686568596  &  24  \\
                 3265651915  &  13  &  1570\phantom{0}  &  2075741463  &  24  &  3938950213  &  24  \\
                 3687027755  &  15  &  1620\phantom{0}  &  2410875726  &  24  &  1183455639  &  24  \\
                 3586546027  &  16  &  1648\phantom{0}  &  3460233636  &  24  &  2419013898  &  24  \\
                 3602222507  &  16  &  1699\phantom{0}  &  4258250079  &  24  &  3864885105  &  24  \\
                 3650858107  &  14  &  1704\phantom{0}  &  3375209167  &  24  &  3796856390  &  24  \\
                 3437182379  &  12  &  1710\phantom{0}  &  1800368255  &  23  &  2177538044  &  24  \\
                 4112721003  &  14  &  1726\phantom{0}  &  3545442555  &  24  &  3432760323  &  24  \\
                 3697088363  &  15  &  1743\phantom{0}  &  1054856781  &  24  &  3645416328  &  24 \\
                 3663407467  &  13  &  1790\phantom{0}  &  3930504896  &  24  &  2046051146  &  24  \\
                 3235068267  &  16  &  1797\phantom{0}  &  3421127945  &  24  &  1653095304  &  23  \\
                 4229866859  &  12  &  1872\phantom{0}  &  2184042351  &  23  &  3065271249  &  24  \\
                 3804923435  &  14  &  1956\phantom{0}  &  1837917379  &  23  &  2079520793  &  23  \\
                 2917649707  &  16  &  1969\phantom{0}  &  1061812134  &  23  &  2096832732  &  24  \\
                 2836121387  &  14  &  1991\phantom{0}  &  410197697   &  23  &  2228488688  &  24  \\
                 2273517239  &  14  &  1995\phantom{0}  &  2148212120  &  23  &  3778174847  &  23  \\  \hline
\end{tabular}

\end{appendix}

\vspace{3ex}
{\footnotesize 2020-2-25}
\end{document}